\documentclass[preprint,12pt]{elsarticle}

\usepackage[utf8]{inputenc}
\usepackage{amssymb}

\usepackage{hyperref}
\hypersetup{
      colorlinks = true,
      citecolor = blue,
      linkcolor = blue,
      urlcolor = blue
}

\journal{}

\begin{document}

\begin{frontmatter}



\title{Non-equilibrium free-energy calculation of phase-boundaries using LAMMPS}


\author[IFGW,CEFET]{Samuel Cajahuaringa \corref{cor1}}
\ead{samuelcm@cefetmg.br}
\author[IFGW,CCES]{Alex Antonelli}

\cortext[cor1]{Corresponding author}

\address[IFGW]{Gleb Wataghin Institute of Physics, University of Campinas, Campinas, São Paulo 13083-859, Brazil}
\address[CCES]{Center for Computing in Engineering $\&$ Sciences, University of Campinas, Campinas, São Paulo 13083-859, Brazil}
\address[CEFET]{Centro Federal de Educação Tecnológica de Minas Gerais, Curvelo, Minas Gerais 35790-000, Brazil}

\begin{abstract}
We present a guide to compute the phase-boundaries of classical systems using a dynamic Clausius-Clapeyron integration (dCCI) method within the \texttt{LAMMPS} (Large-scale Atomic/Molecular Massively Parallel Simulator) code. The advantage of the dCCI method is because it provides coexistence curves spanning a wide range of thermodynamic states using relatively short single non-equilibrium simulations. We describe the state-of-the-art of non-equilibrium free-energy methods that allow us to compute the Gibbs free-energy in a wide interval of pressure and/or temperature. We present the dCCI method in details, discuss its implementation in the \texttt{LAMMPS} package and make available source code, scripts, as well as auxiliary files. As an illustrative example, we determine the phase diagram of silicon in a range of pressures covering from 0 to 15 GPa and temperatures as low as 400 K up to the liquid phase, in order to obtain the phase boundaries and triple point between diamond, liquid and $\beta$-Sn phases.
\end{abstract}

\begin{graphicalabstract}
\end{graphicalabstract}

\begin{highlights}
\item Research highlight 1
\item Research highlight 2
\end{highlights}


\end{frontmatter}


\section{Introduction}
%

The knowledge of phase diagrams is of great importance for many research areas in physics, chemistry, and engineering. A phase diagram is a map used to show in which conditions of pressure, temperature, volume, etc, one can find each specific phase of a system and the boundaries between these various phases.

The determination of phase diagrams by computer simulation requires the computation of the free-energy of the various phases of the system, this task is far from trivial. The difficulties stem from the fact that entropy and free-energies depend on the volume in phase space available to the system, in contrast to other thermodynamic properties, such as internal energy, enthalpy, temperature, pressure, etc., that can be easily computed as simple averages of functions of the phase space coordinates.

The determination of phase diagrams by computer simulation can be achieved by employing free-energy evaluation techniques such as thermodynamic integration (TI) \cite{Kirkwood1935}, which consists in the construction of a sequence of equilibrium states along a path between two thermodynamic states of interest.
The calculated free-energy difference between these two states is essentially exact, if the free-energy of one of them is known, which we call reference system. From that the absolute free-energy of the other system, which we call system of interest, is readily obtained. In practice, since the integration along the thermodynamic path is performed numerically, it requires the calculation of several ensemble averages, which is computationally very demanding.

In recent years, several studies\cite{Freitas2016,Leite2016,Leite2019,Cajahuaringa2018,Cajahuaringa2019} have been proving the efficiency of the non-equilibrium (NE) techniques, making free-energy  calculations much more accessible\cite{Freitas2016,Leite2019}, which were made available in the \texttt{LAMMPS} molecular dynamics (MD) package\cite{LAMMPS}. In contrast to the equilibrium TI method, the NE approaches estimate the desired free-energy difference by traversing the thermodynamic path between the systems of interest and reference in an explicitly time-dependent process, which have been shown to give accurate results using only a few relatively short non-equilibrium simulations. 

However, several NE calculations are required for each coexistence condition between the phases of the system to map correctly the phase boundaries. In order to optimize the calculation of phase boundaries, it was proposed by de Koning et al.\cite{deKoning2001} a method for the integration of the Clausius-Clapeyron equation using non-equilibrium simulations. This method allows the calculation of the phase boundary using, in principle, a single simulation, whose length is comparable to a regular equilibrium simulation.

In this paper, we demonstrate the potential of the dynamic Clausius-Clapeyron integration (dCCI) method by using it to obtain, entirely from atomistic simulations employing a realistic interatomic potential, the phase diagram in a wide range of temperatures and pressures, focusing on the implementation in the widely used \texttt{LAMMPS} MD package. We show how the NE methods can be used to compute the Gibbs free-energy of different phases, in a broad range of temperature and/or pressure, in order to determine an initial coexistence condition and, from that, how the dCCI method can be used to accurately and efficiently compute the phase boundaries of atomistic systems. We provide excerpts from the used \texttt{LAMMPS} scripts to exemplify the practical details of the calculations. Complete \texttt{LAMMPS} scripts, source codes and post-processing tools have also been made available in the following github account. As an illustration we performed non-equilibrium free-energy calculations of the phase-boundaries of silicon, described by the Stillinger-Weber\cite{Stillinger1985} potential.

The paper is organized as follows. In Sec.~\ref{sec:NFEM} we review the relevant aspects of NE free-energy methods to calculate the Gibbs free-energy along isothermal and isobaric paths. Sec. \ref{sec:DCCI} provides the details of the dCCI methodology, in which from a given initial coexistence condition one is able to determine the entire phase boundary from a single non-equilibrium simulation. Sec. \ref{sec:applications} describes the application of dCCI to compute the phase diagram of Si. We end with a summary and conclusions in Sec. \ref{sec:conclusions}.
\section{Non-equilibrium free-energy methods}
\label{sec:NFEM}
The TI method\cite{Kirkwood1935} has been widely used to compute Helmholtz free-energy differences by using the calculation of reversible work. It consists in the construction of a path connecting two thermodynamic states by defining a parametrized Hamiltonian $H(\lambda)$, where $\lambda$ is a coupling parameter describing the interpolation between the two ends of the thermodynamic path.

It is straightforward to show, starting from the definition of the partition function, that the Helmholtz free-energy difference between the two equilibrium states corresponding to $H(\lambda_{i})$ and $H(\lambda_{f})$ is given by the reversible work $W_{rev}$ done along the quasi-static process that connects these states, i.e.
\begin{equation}
\Delta F = F(\lambda_{f})-F(\lambda_{i})=\int_{\lambda_{i}}^{\lambda_{f}}\left \langle\frac{\partial H}{\partial \lambda}\right \rangle d\lambda\equiv W_{rev},
\label{eqn:TI}
\end{equation}
where $\left\langle\cdots \right\rangle_{\lambda}$ is the canonical ensemble average at a particular value of parameter and $\frac{\partial H}{\partial \lambda}$ is the so-called driving-force. In this approach, the integral is discretized on a grid of $\lambda$-values between $\lambda_{i}$ and $\lambda_{f}$ and a separate equilibrium simulation is performed for each value of lambda.

In the NE approach, the adiabatic switching\cite{Watanabe1990} (AS) method estimates the integral in Eq.~\ref{eqn:TI} by replacing the integration over equilibrium ensemble averages, by an integration over instantaneous values of driving force along a single simulation in which $\lambda=\lambda(t)$ changes continuously throughout the simulation 
\begin{equation}
W_{dyn}=\int_{0}^{t_{s}}\frac{d\lambda}{dt}\frac{\partial H(\lambda)}{\partial \lambda}\Bigg|_{\lambda(t)} dt,
\label{eqn:Wdyn}
\end{equation}
where $t_{s}$ is the total time of the switching process and $W_{dyn}$ is the dynamical work done. Due to the irreversible nature of the process, entropy is produced, causing $W_{dyn}$ to be a stochastic variable whose mean value, by the second law of thermodynamics, differs from $W_{rev}$ according to the relation
\begin{equation}
\Delta F=W_{rev}=\overline{W}_{dyn}-\overline{Q}_{diss},
\end{equation}
where the overbar means an average over an ensemble of realizations of the AS process and $\overline{Q}_{diss}\geqslant 0$ is the average dissipated heat, which is zero only in the quasi-static limit ($t_{s}\to \infty$). However, provided that the non-equilibrium process is sufficiently ''close'' to the ideal quasi-static process, hence the linear response theory is valid, it can be shown that the systematic error can be eliminated by combining the results of the processes realized in both directions (forward and backward). Accordingly, one has
\begin{equation}
\Delta F= \frac{1}{2}[\overline{W}_{dyn}^{i\to f}-\overline{W}_{dyn}^{f\to i}],
\label{eqn:dF}
\end{equation}
where we have used the fact that within the linear response theory the heat dissipated in both processes are identical $\overline{Q}_{diss}^{i\to f}=\overline{Q}_{diss}^{f\to i}$. Similarly, the dissipated heat can be estimated by
\begin{equation}
\overline{Q}_{diss}^{i\to f}=\overline{Q}_{diss}^{f\to i}=\frac{1}{2}[\overline{W}_{dyn}^{i\to f}+\overline{W}_{dyn}^{f\to i}],
\label{eqn:Qdiss}
\end{equation}
Eqs.~\ref{eqn:dF} and \ref{eqn:Qdiss} can be used for monitoring of the convergence of $\Delta F$ and the heat dissipation as a function of the switching time $t_{s}$. The AS method can lead to significant efficiency gains when compared to standard equilibrium methods, providing accurate estimates of $\Delta F$ using only a few relatively short non-equilibrium simulations.

Next we discuss the application of three specific thermodynamic paths for $H(\lambda)$ that allow the calculation of, (i) the Helmholtz free-energy difference between two systems described by different Hamiltonians, (ii) the temperature-dependence of the Gibbs free-energy for a given system Hamiltonian along an isobaric path, and (iii) the pressure-dependence of the Gibbs free-energy for a given system Hamiltonian along an isothermal path.
\subsection{Helmholtz free-energy difference between two systems: Hamiltonian interpolation
method}\label{subsec:AS}
Suppose we wish to compute the Helmholtz free-energy of some system of interest described by a Hamiltonian $H_{int}$ of the form
\begin{equation}
    H_{int}=K+U_{int}(\textbf{r}),
\end{equation}
where $K$ is the kinetic energy and $U_{int}(\textbf{r})$ is the system of interest’s potential energy, where $\textbf{r}$ stands for the set of particles coordinates. A second system, in the same thermodynamic phase of the system of interest, can be used as a reference system if its Helmholtz free-energy is known (analytically or numerically), described by the Hamiltonian $H_{ref}$, given by
\begin{equation}
    H_{ref}=K+U_{ref}(\textbf{r}_{i}),
\end{equation}
with $U_{ref}(\textbf{r})$ being its potential energy.

A typical functional form of $H(\lambda)$ that couples the systems of interest and reference is given by a linear interpolation as follow
\begin{equation}
    H_{\lambda}=\lambda H_{int}+(1-\lambda)H_{ref},
\end{equation}
note that this form allows a continuous switching between $H_{int}$ and $H_{ref}$ by varying the coupling parameter $\lambda$ between $\lambda_{i}=1$ and $\lambda_{f}=0$. According to Eq. \ref{eqn:TI}, the reversible work is given by 
\begin{equation}
W_{rev}=\int_{\lambda_{i}}^{\lambda_{f}}\langle U_{int}-U_{ref}\rangle d\lambda
\label{eqn:Wrev_HI}
\end{equation}
and its corresponding dynamical work estimator for the forward process is given by
\begin{equation}
W_{dyn}^{i\to f}=\int_{0}^{t_{s}}\frac{d\lambda}{dt}(U_{int}-U_{ref})dt,
\end{equation}
note that for the backward process the time derivative $\frac{d\lambda}{dt}$ has the reversed sign of the forward process. By combining the average results obtained from a number of independent realizations of both switching processes, the desired Helmholtz free-energy is estimated as
\begin{equation}
F_{int}=F_{ref}+\frac{1}{2}[\overline{W}_{dyn}^{i\to f}-\overline{W}_{dyn}^{f\to i}].
\label{eqn:dA}
\end{equation}
For solids systems, either crystalline or amorphous, the choice of the reference system is straightforward, the Einstein crystal\cite{Frenkel,Freitas2016,deKoning1997} whose Helmholtz free-energy is analytically known. For fluid-phase systems recently it was proposed that the Uhlenbeck-Ford (UF) model\cite{Leite2016,Leite2019}, which is a robust reference system that provides an accurate fluid-phase Helmholtz free-energy.
\subsection{Gibbs free-energy as a function of the temperature: the Reversible Scaling method}
Suppose the Gibbs free-energy $G_{int}(P,T_{0})$ of a system of interest is known at some temperature $T_{0}$ and fixed pressure $P$, and we now wish to determine $G_{int}(P,T)$ for another temperature $T$ along an isobaric path. That can be achieved in a single simulation by using the reversible scaling (RS) technique\cite{deKoning1997}. The RS method in the NPT ensemble\cite{deKoning2001} is based on the following parametric Hamiltonian
\begin{equation}
H_{RS}(\lambda)=K+\lambda U_{int}(\textbf{r})+P_{S}(\lambda)V,
\label{eqn:H_RS}
\end{equation}
in this case, this Hamiltonian represents the enthalpy of the scaled system, in which the potential energy is scaled by the coupling parameter and $P_{S}$ the external pressure on the scaled system. The configurational and volume contributions to the partition function in the NPT ensemble $Z_{RS}(P_{S}(\lambda),\lambda)$ for the RS Hamiltonian at temperature $T_{0}$ and pressure $P_{s}$ is given by
%
\begin{eqnarray}
Z_{RS}(P_{S}(\lambda),\lambda) &=& \int dV\exp[-P_{S}(\lambda)V/k_{B}T_{0}]\int_{V}d^{3N}\mathbf{r}\exp[-\lambda U_{int}(\textbf{r})/k_{B}T_{0}] \nonumber \\
&=& Z_{int}(P,T),
\label{eqn:Z_RS}
\end{eqnarray}
%
which is equal to the partition function of the system of interest at temperature $T$ and pressure $P$ by using the following scaling relations:
\begin{equation}
T=\frac{T_{0}}{\lambda}
\label{eqn:T_sc}
\end{equation}
and
\begin{equation}
P_{S}(\lambda)=\lambda P.
\label{eqn:P_sc}
\end{equation}
On the basis of Eqs. \ref{eqn:Z_RS}, \ref{eqn:T_sc}, and \ref{eqn:P_sc}, the Gibbs free energies of the system of interest $G_{int}$ and the scaled system $G_{RS}$, are related according to
\begin{equation}
G_{int}(P,T)=\frac{G_{RS}(P_{S}(\lambda),T_{0};\lambda)}{\lambda}+\frac{3}{2}Nk_{B}T_{0}\frac{\ln\lambda}{\lambda}
\label{eq:GRS}
\end{equation}
where $N$ is the number of atoms. Eq.~\ref{eq:GRS}, implies that each value of $\lambda$ in the scaled Hamiltonian $H_{RS}$ at a fixed temperature $T_{0}$ and scaled pressure $P_{S}(\lambda)$ corresponds to the system of interest described by $H_{int}$ at a temperature $T$ and pressure $P$. Thus, the task of calculating the free-energy of the physical system as a function of temperature $T$ at fixed pressure $P$ can be accomplished from $H_{RS}(\lambda)$ by varying the scaling parameter $\lambda$ and the scaling pressure $P_{S}(\lambda)$ at fixed temperature $T_{0}$. In this case one can identify  the driving force as
\begin{equation}
    \frac{\partial H_{RS}(\lambda)}{\partial \lambda}=U_{int}(\textbf{r})+\frac{dP_{S}(\lambda)}{d\lambda}V.
    \label{eqn:dH_RS}
\end{equation}
The reversible work in the RS method is given by
\begin{equation}
    W_{rev}(\lambda)=\int_{1}^{\lambda}\Bigg(\langle U_{int}\rangle+\frac{dP_{S}(\lambda')}{d\lambda'}\langle V\rangle\Bigg)d\lambda',
    \label{eqn:Wrev_RS}
\end{equation}
can be obtained within very good accuracy through the AS method, which is used to estimate the forward dynamical work, with $\lambda(t)$ varying from $\lambda(0)=1$ to $\lambda(t_{s})=\lambda_{f}$
\begin{equation}
    W_{dyn}^{i\to f}=\int_{0}^{t_{s}}\frac{d\lambda}{dt}\Bigg(U_{int}+\frac{dP_{S}(\lambda)}{d\lambda}\Bigg|_{\lambda(t)}\Bigg)dt.    
    \label{eqn:Wdyn_RS}
\end{equation}
By carrying out the scaling process in the opposite direction and using Eq.~\ref{eqn:dF}, $G_{int}(P,T)$ is given by
%
\begin{equation}
G_{int}(P,T)=\frac{G_{int}(P,T_{0})}{\lambda}+\frac{3}{2}Nk_{B}T_{0}\frac{\ln\lambda}{\lambda}+\frac{1}{2\lambda}[\overline{W}_{dyn}^{1\to \lambda}-\overline{W}_{dyn}^{\lambda\to 1}],
\label{eqn:Gibbs_RS}
\end{equation}
where $\lambda$ varies between $1$ and $\lambda_{f}$, and $G_{int}(P,T_{0}) = G_{RS}(P,T_{0})$.

It is important to note that the application of this approach requires the knowledge of the absolute Gibbs free-energy of the system of interest at a temperature $T_{0}$ and pressure $P$, which can be computed by knowing the Helmholtz free-energy at temperature $T_{0}$ and the average volume of the system $V$, which corresponds to the external pressure $P$, through the thermodynamic relation
\begin{equation}
G_{int}(P,T_{0})=F_{int}(V,T_{0})+PV
\label{eqn:G_and_A}
\end{equation}
where the $F(T_{0},V)$ can be determined by employing the AS method, detailed in \ref{subsec:AS}.
\subsection{Gibbs free-energy as a function of pressure: Adiabatic Switching method}
Suppose the Gibbs free-energy $G_{int}(P_{0},T)$ of a system of interest is known at some pressure $P_{0}$ and fixed temperature $T$, and we now wish to determine $G_{int}(P,T)$ for other pressure $P$ along an isothermal path, where the Hamiltonian of the interest system is 
\begin{equation}
H_{int}=K+U_{int}(\textbf{r})+PV,    
\end{equation}
if we take the external pressure as coupling parameter, the work associated with a reversible process along a path connecting the physical system at the reference pressure $P_{0}$ to the system at a pressure of interest $P$ is
\begin{equation}
G_{int}(P,T)-G_{int}(P_{0},T)=\int_{P_{0}}^{P}\left \langle\frac{\partial H}{\partial P'}\right \rangle dP'=\int_{P_{0}}^{P}\langle V\rangle dP'\equiv W_{rev},
\label{eqn:Wmech}
\end{equation}
where the brackets in Eq.~\ref{eqn:Wmech} indicate the equilibrium average of the volume in the isothermal-isobaric ensemble. An efficient alternative to performing several equilibrium simulations can be achieved by using the AS method\cite{Cajahuaringa2018}, the integral can be evaluated along a single non-equilibrium simulation during which the magnitude of the external pressure $P(t)$ changes dynamically, in such a way that, at the beginning of the simulation $P(0)=P_{0}$ and at the end $P(t_{s})=P$. In this case the reversible work will be estimated in terms of the dynamical work 
\begin{equation}
W_{dyn}=\int_{0}^{t_{s}}\frac{dP(t)}{dt}\bigg|_{t}V(t)dt.
\label{eqn:Wmech_dyn}
\end{equation}
In this switching process not only the initial and final points on the trajectory correspond to physical systems, as it is usual in AS simulations, but the information gathered at the intermediate states of the path also has physical meaning. As a consequence, one obtains the Gibbs free-energy of a system in a wide interval of pressure, provided that the Gibbs free-energy of the system of interest in a reference state is known.
\begin{equation}
G_{int}(P,T)=G_{int}(P_{0},T)+\frac{1}{2}[\overline{W}_{dyn}^{i\to f}-\overline{W}_{dyn}^{f\to i}].
\label{eqn:Gibbs_AS}
\end{equation}
\section{Dynamic Clausius-Clapeyron integration}\label{sec:DCCI}
First-order phase transitions for pure substances, are usually represented as lines in a PT diagram, which indicate the phase boundaries between the phases and its slope can be described by the Clausius–Clapeyron equation (CCE), which uses thermodynamic properties of the both coexisting phases. The CCE in the PT diagram can be written in several forms, the most common one is,
\begin{equation}
    \frac{dP}{dT}=\frac{\Delta H}{T\Delta V},
    \label{eqn:CCE}
\end{equation}
where $\Delta H$ and $\Delta V$ are the molar enthalpy and molar volume differences between the two phases, respectively. The CCE is remarkable because if a point of the coexistence curve is known, one can obtain the whole phase boundary from Eq. \ref{eqn:CCE}.

In the nineties Kofke\cite{kofke1993} proposed a method that employs the CCE combined with results from computer simulations to calculate phase boundaries. The method is very effective, however, it requires a series of independent equilibrium simulations of both phases of interest, which is quite demanding in terms of computing time.

In order to avoid performing a series of independent equilibrium simulations for each phase required to map the phase boundaries, de Koning et al.\cite{deKoning2001} based on a similar idea to the RS method, proposed a method in which the integration of Clausius-Clapeyron equation is performed dynamically. Starting at a single point of the coexistence curve, this technique allows the exploration of the coexistence curve over a wide range of states from two non-equilibrium simulations, one for each phase, running simultaneously.

Let us start from a known condition of phase coexistence of two phases $I$ and $II$, that occurs when the Gibbs free energies of both phases are equal $G_{s,I}=G_{s,II}$, if reversible perturbations $d\lambda$ and $d\lambda(dP_{S}/d\lambda)$ are applied, the Gibbs free energies of the scaled systems will change according to
\begin{equation}
dG_{S,I}=dW_{rev,I}=d\lambda\Bigg[\left \langle U_{int}\right \rangle_{I}+\frac{dP_{S}}{d\lambda}\left \langle V\right \rangle_{I}\Bigg]
\end{equation}
and
\begin{equation}
dG_{S,II}=dW_{rev,II}=d\lambda\Bigg[\left \langle U_{int}\right \rangle_{II}+\frac{dP_{S}}{d\lambda}\left \langle V\right \rangle_{II}\Bigg],
\end{equation}
where all ensemble averages are evaluated at temperature $T_{0}$, scaling parameter $\lambda$, and scaled pressure $P_{S}(\lambda)$. In order to maintain phase-coexistence upon the application of this disturbance, one must have that $dG_{S,I}= dG_{S,II}$, from this condition we obtain the following relationship between pressure and temperature of the scaled systems,
\begin{equation}
\frac{dP_{S}}{d\lambda}=-\Bigg(\frac{\left \langle U_{int}\right \rangle_{I}-\left \langle U_{int}\right \rangle_{II}}{\left \langle V\right \rangle_{I}-\left \langle V\right \rangle_{II}}\Bigg),
\label{eqn:RS_CCE}
\end{equation}
this equation is the CCE for the scaled systems, where the temperature is represented by the coupling parameter $\lambda$.

Similarly to what is done to estimate the free-energy from non-equilibrium simulations, we can modify Eq.~\ref{eqn:RS_CCE} in a way that the parameter $\lambda$ is varied dynamically, transforming it into the dynamic Clausius-Clapeyron equation 
\begin{equation}
\frac{dP_{S}}{dt}=-\frac{d\lambda}{dt}\Bigg(\frac{U_{int,I}(t)-U_{int,II}(t)}{V_{I}(t)-V_{II}(t)}\Bigg),
\end{equation}
where the ensemble averages of the potential energies and volumes have been replaced by instantaneous values along the time-dependent process. Provided that the dynamic process is ideally reversible, the coexistence curve is given by
\begin{equation}
P_{S,coex}(\lambda(t_{s}))=P_{S,coex}(\lambda=1)-\int_{0}^{t_{s}}\frac{d\lambda}{dt}\Bigg(\frac{U_{int,I}(t)-U_{int,II}(t)}{V_{I}(t)-V_{II}(t)}\Bigg)dt.
\label{eqn:Ps_integration}
\end{equation}

Given an initial coexistence condition, the integration of this equation then provides a dynamic estimator for the entire coexistence curve from two non-equilibrium simulations, one for each phase, which are performed simultaneously. In practice, one ought to consider finite changes of the parameter $\lambda$ in Eq.~\ref{eqn:Ps_integration}, in such a way we obtain
\begin{equation}
P_{S,coex}(\lambda_{k+1})=P_{S,coex}(\lambda_{k})-(\lambda_{k+1}-\lambda_{k})\Bigg(\frac{\Delta U_{I-II,k}}{\Delta V_{I-II,k}}\Bigg)
\label{eqn:Ps_coex_discreted}
\end{equation}
where $\Delta U_{I-II,k}=U_{int,I}(\textbf{z}_{k})-U_{int,II}(\textbf{z}_{k})$ and $\Delta V_{I-II,k}=V_{I,k}-V_{II,k}$, in which $\textbf{z}_k$ stands for the phase space coordinates of the $k$-th microstate. The initial value of $\lambda$ is always set to be equal to 1, and it is varied according to Eq.~\ref{eqn:T_sc} in order to attain the desired temperature $T$, depending on the rate of change of $\lambda$ a given scaled coexistence pressure $P_{S,coex}(\lambda)$ can be attained for different values of $\lambda$ from Eq.~\ref{eqn:Ps_coex_discreted}. Using the scaling relation given by Eq.~\ref{eqn:P_sc}, we obtain the coexistence pressure $P_{coex}$, in other words, we obtain the coexistence curve as the pressure coexistence, $P_{coex}(T)$, for a wide interval of the temperature.

If the coexistence curve has a small slope (in absolute value) i.e. the pressure derivative with respect the temperature has a small value, which is common in solid-solid coexistence lines, it is more convenient use the $\lambda$ parameter as control parameter and the Eq.~\ref{eqn:Ps_coex_discreted}. If, on the other hand, the coexistence curve is expected to have a large slope (in absolute value) i.e. the pressure derivative with respect the temperature has a large value, which is common in solid-fluid and fluid-fluid phase boundaries, in these case is more convenient use the pressure $P$ as control parameter, which can be obtained by inverting the relationship between $\lambda$ and $P_{S,coex}$ in Eq.~\ref{eqn:Ps_coex_discreted}
\begin{equation}
\lambda_{k+1}=\lambda_{k}\Bigg(\frac{1+P_{k}\Delta V_{I-II,k}/\Delta U_{I-II,k}}{1+P_{k+1}\Delta V_{I-II,k}/\Delta U_{I-II,k}}\Bigg)
\label{eqn:l_coex_discreted}
\end{equation}
and by using the scaling relations Eqs.~\ref{eqn:P_sc} and ~\ref{eqn:T_sc}, we obtain the  coexistence curve as the coexistence temperature in a wide interval of pressure $T_{coex}(P)$.

Evidently, the dynamic estimators of $P_{coex}(T_{k})$ or $T_{coex}(P_{k})$ are subject to errors associated with the irreversible nature of the time-dependent scaling process, however those effects decrease quickly with increasing number of simulation steps, allowing an accurate determination of the entire phase boundary from a couple of single, relatively short dynamic scaling simulations.
\section{Application: results and discussion}\label{sec:applications}
In this section we describe the implementation of the dCCI method in the widely used Molecular Dynamics code \texttt{LAMMPS}. We demonstrate the use of the NE methods to calculate coexistence conditions by computing the Gibbs free-energy of the phases in a wide interval of temperature and pressure. Through the use of the dCCI method we compute the silicon phase boundaries, described by the Stillinger-Weber (SW)\cite{Stillinger1985} potential. Complete \texttt{LAMMPS} scripts, source code and auxiliary files are available on the following github project \cite{Samuelgithub}.

Initially we need to determine the phase coexistence conditions: (i) the melting temperature of silicon in the diamond phase (Si-cd) at 0.0 GPa. Next, we compute (ii) the melting temperature of silicon in the $\beta$-tin phase (Si-$\beta$-tin) at 15.0 GPa. Then, we compute (iii) the coexistence pressure between Si-cd and Si-$\beta$-tin at 400 K. Finally, using these coexistence conditions (iv) we apply the dCCI method to calculate the phase diagram of silicon and determine the triple point predicted by the SW potential.
\subsection{Melting points of silicon}
First, we determine the melting points of Si-cd phase at zero pressure and Si-$\beta$-tin at $15.0$ GPa. To this end, we need compute the Gibbs free-energy curves $G(P,T)$ for the solids and liquid phases in order to determine the melting temperature $T_{m}$ at their crossing points. To achieve this we follow Refs. \citenum{Freitas2016} and \citenum{Leite2019}, which provides details of the methodologies for calculating the free-energy of the solid and liquid phases using the \texttt{LAMMPS} package.

We use computational cells containing 8000 atoms in the calculations of the Si-cd and liquid phases, and in the case of Si-$\beta$-tin it was used a cell of 7448 atoms. All systems were subject to periodic boundary conditions. Pressure and temperature control was attained using a Parrinello-Rahman\cite{Parrinello1981} barostat and a Langevin\cite{Schneider1978} thermostat. The corresponding equations of motion were integrated using the velocity-Verlet algorithm with a time step of $\Delta t= 1$ fs.

In order to compute the Gibbs free-energy of Si-cd as a function of the temperature at zero-pressure, first we determine the equilibrium volume at $T_{0}=400$ K, using an initial equilibration of 0.2 ns and then averaging the volume over a time interval of 1.0 ns. Next, we compute the Gibbs free-energy at the reference temperature of 400 K, using as reference system the Einstein crystal with a spring constant of 6.113 eV/\AA$^{2}$. The system was then equilibrated during 0.1 ns prior to the switching runs. The unbiased dynamical-work estimators were obtained from 10 independent forward and backward realizations using a switching time of 0.2 ns to converge the results. Finally, we use the RS method to compute the Gibbs free-energy curve of Si-cd as a function of temperature at 0.0 GPa. The RS path was chosen in such a way that the scaling parameter $\lambda$ was varied between $\lambda=1$ to $\lambda=\lambda_{f}=400/2000$, thus, covering a temperature range between $T_{0}=400$ K and $T_{f}=2000$ K, within a switching time of $t_{s}=0.5$ ns. The dynamical work was determined as the average over 10 independent realizations of the RS process in the forward and backward directions. Full details of the methodology to compute the Gibbs free-energy curve as a function of temperature at zero pressure are described in Ref. \citenum{Freitas2016}.
\begin{figure}[tbp]
    \centering
         \includegraphics[scale=0.44]{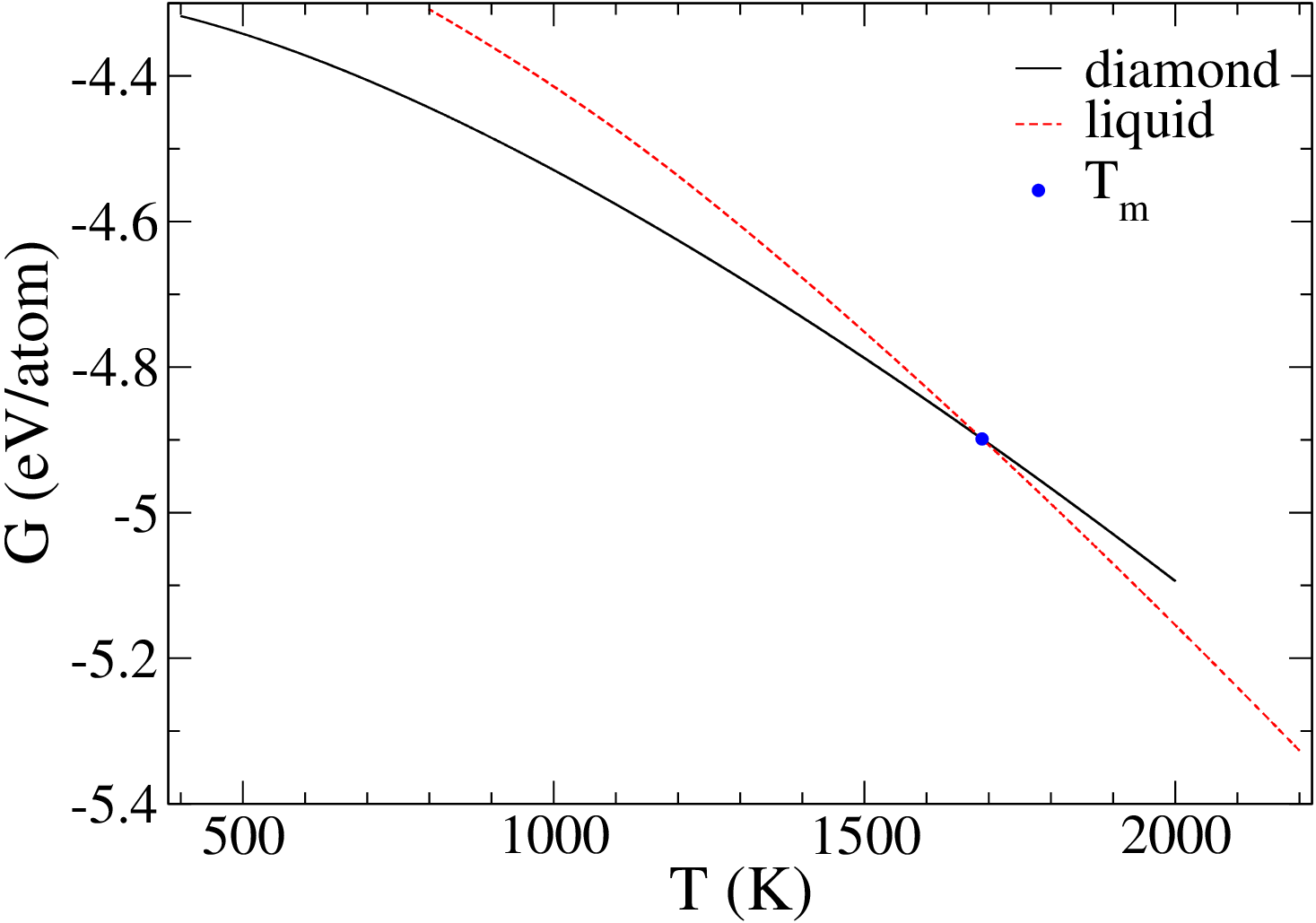}
     \caption{\label{fig:Tm1} Gibbs free-energy per atom of Si-cd and liquid phases at zero pressure. The crossing of the curves indicates the melting temperature at zero pressure.}
\end{figure}

Now we compute the absolute Gibbs free-energy of liquid phase at zero-pressure as a function of the temperature. We start by determining the equilibrium volume at $T_{2}=2200$ K. After an initial equilibration of 0.2 ns, the average value of the volume is determined over a time interval of 1.0 ns. Next, we compute the Gibbs free-energy at a reference temperature 2200 K, using as reference system the UF model\cite{Leite2016}, with the following parameters: p=25 and $\sigma=2.0$. The system was equilibrated during 0.1 ns prior to the switching runs. The unbiased dynamical-work estimators were obtained from 10 independent forward and backward realizations using a switching time of 0.1 ns to converge the results. Finally, we use the RS method to compute the Gibbs free-energy curve of the liquid phase as a function of temperature at 0.0 GPa. In this case, we used a RS path in which the scaling parameter $\lambda$ was varied between $\lambda=1$ to $\lambda=\lambda_{f}=2200/800$, thus covering a temperature range between $T_{0}=2200$ K and $T_{f}=800$ K, within a switching time of $t_{s}=0.5$ ns. The dynamical work was determined as the average over 10 independent realizations of the RS process along the forward and backward directions. Full details of the calculation of the Gibbs free-energy curve of the liquid phase as a function of temperature at zero pressure are described in Ref.~\citenum{Leite2019}. Fig. \ref{fig:Tm1} shows the melting temperature of Si-cd at 0.0 GPa at the value of $T_{m}=1689.2(3)$ K. This result is in agreement with that previously reported in Ref. \citenum{Ryu2008}.

\begin{figure}[tbp]
    \centering
         \includegraphics[scale=0.44]{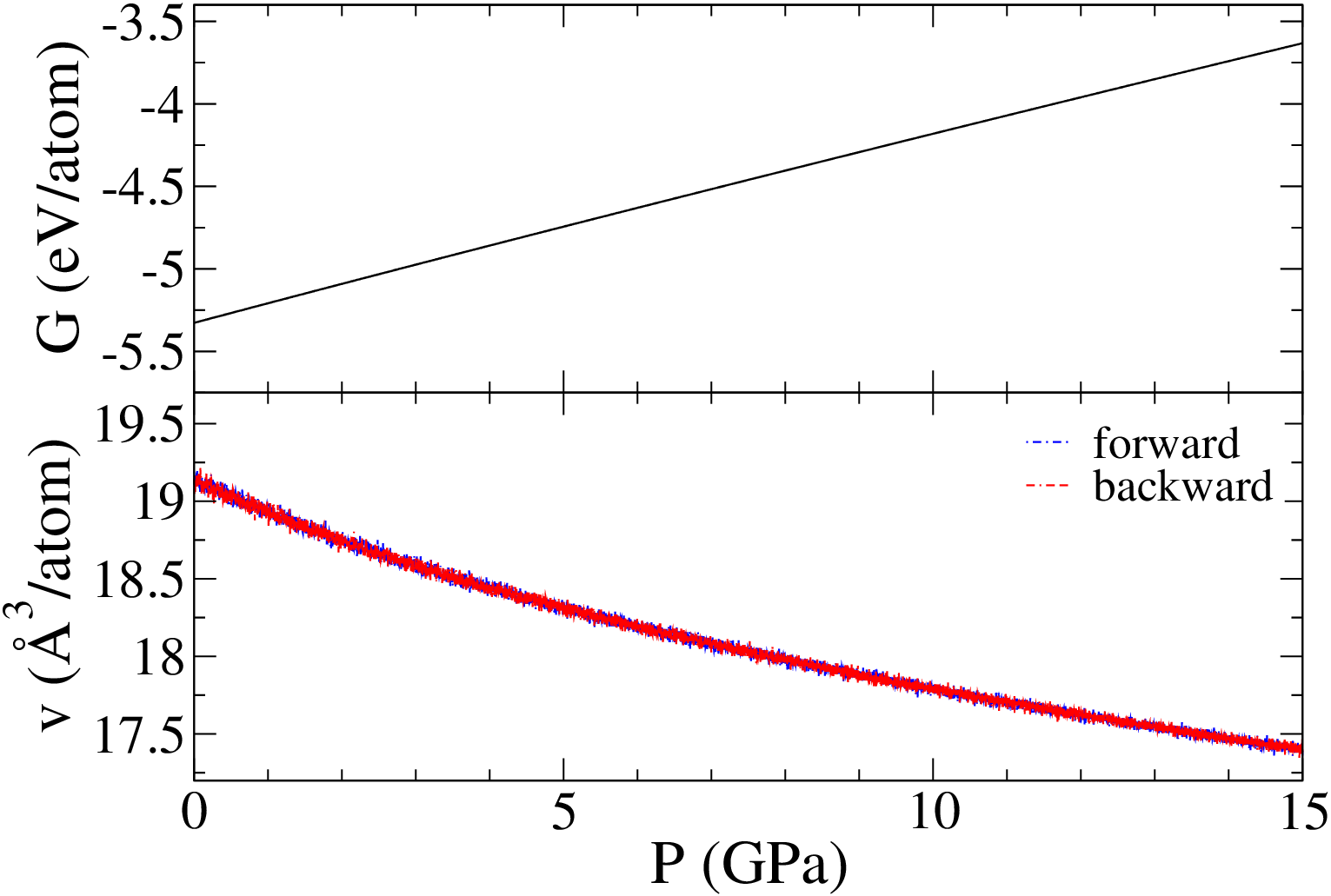}
     \caption{\label{Fig:AS1}(\textbf{Upper panel}) Gibbs free-energy of liquid Si as a function of pressure at 2200~K. \textbf{Lower panel} volume per atom as a function of the dynamical pressure along the forward and backward processes}
\end{figure}

After computing the melting point at zero pressure, we now determine the second melting of silicon at 15.0 GPa by the crossing of the absolute Gibbs free-energy curves of the Si-$\beta$-tin and liquid phases. This is accomplished using a similar procedure used for Si-cd. To compute the absolute Gibbs free-energy as a function of temperature for the Si-$\beta$-tin phase, we determine the equilibrium volume at $T_{1}=400$ K, using an initial equilibration of 0.2 ns and then the average value of the volume is determined over a time interval of 1.0 ns. Next, we compute the Helmholtz free-energy at the temperature of 400 K (Eq.~\ref{eqn:dA}), using as reference system the Einstein crystal with a spring constant of 1.362 eV\AA$^{2}$, the system was equilibrated during 0.1 ns prior to the switching runs. The unbiased dynamical-work estimators were obtained from 10 independent forward and backward realizations using an switching time of 0.5 ns. From these calculations, the Gibbs free-energy is obtained from the Eq.~\ref{eqn:G_and_A}. Finally, we use the RS method to compute the Gibbs free-energy curve as a function of temperature at 15.0 GPa. In this case, during the RS path, aside from the scaling of the force done by the \texttt{fix adapt} command in LAMMPS, we need to scale the external pressure of the barostat as well, in order to take into account the effect of volume change of the system. To this end, we modified the \texttt{fix npt (fix nph)} to include a new tag \texttt{ners}, which multiply the external pressure by the scaling parameter $\lambda$. This is achieved using the following fix commands in the \texttt{LAMMPS} script:

\texttt{variable lambda equal 1/(1+(elapsed/\$\{t\_s\})*(1/\$\{lf\}-1))}

\texttt{fix f1 all nph aniso 15000.0 15000.0 1.0 ners v$\_$lambda}

The RS path chosen was such that the scaling parameter $\lambda$ is varied between $\lambda=1$ to $\lambda=\lambda_{f}=400/1600$. Thus, covering the temperature range between $T_{0}=400$ K and $T_{f}=1600$ K. The temperature dependence of the Gibbs free-energy is then computed using Eq.~\ref{eqn:Gibbs_RS}, with a switching time of $t_{s}=0.5$ ns. The dynamical work was determined as the average over 10 independent realizations of the RS process along the forward and backward directions.

\begin{figure}[tbp]
    \centering
         \includegraphics[scale=0.44]{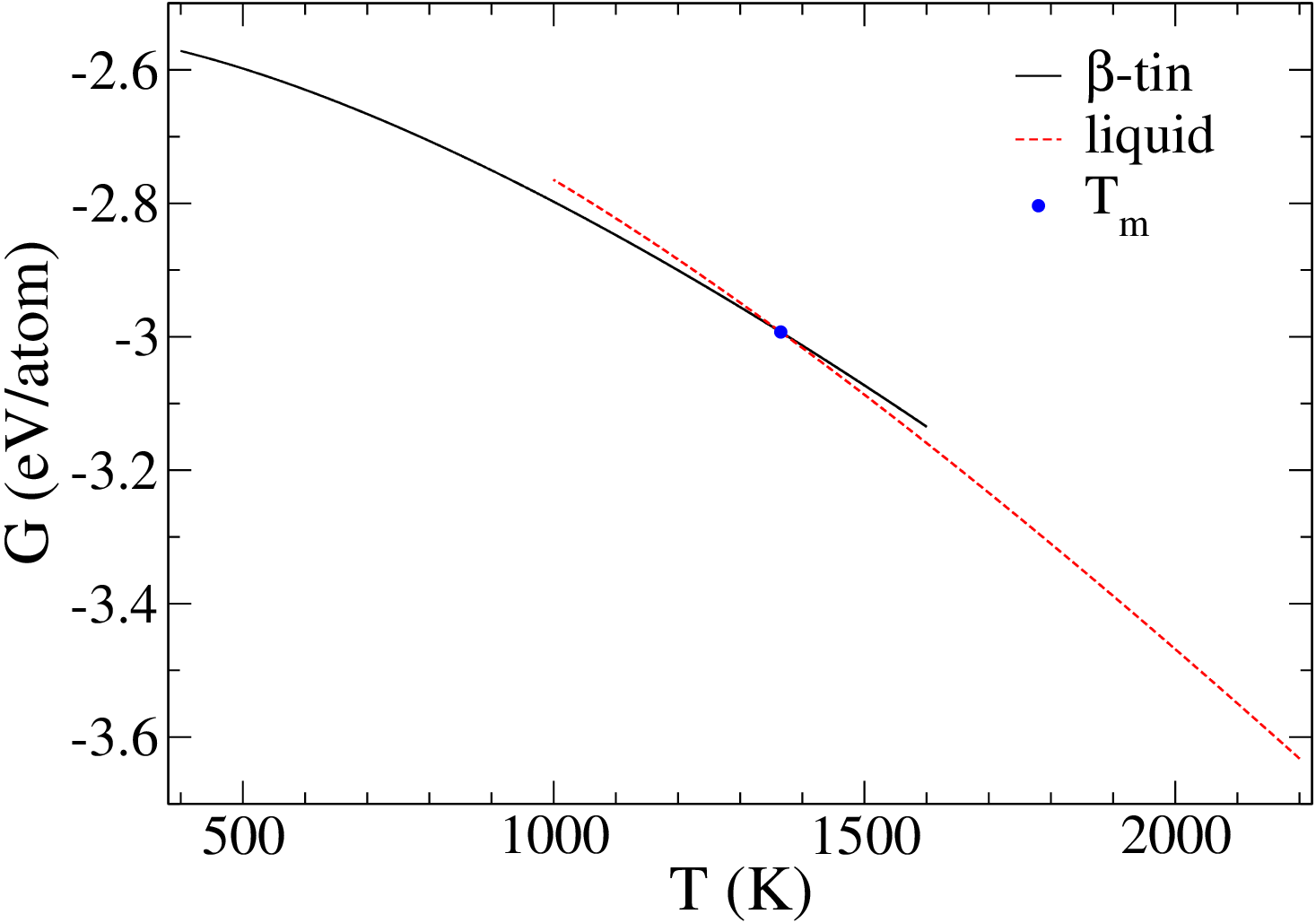}
     \caption{\label{fig:Tm2} Gibbs free-energy per atom of Si-$\beta$-tin and liquid phases at 15.0 GPa. The crossing of the curves indicates the melting temperature at 15.0 GPa.}
\end{figure}

To compute the absolute Gibbs free-energy as a function of the temperature of the liquid phase at 15.0 GPa, we follow a different procedure. First, we determine the Gibbs free-energy as a function of pressure at 2200 K using the AS method (Eq.~\ref{eqn:Gibbs_AS}), in which the external pressure $P$ is changed dynamically from 0.0 GPa to 15.0 GPa. The system was equilibrated during 0.1 ns prior to the switching runs, and the unbiased dynamical work (Eq.~\ref{eqn:Wmech_dyn}) was obtained from 10 independent forward and backward realizations using switching times between 0.01 ns to 0.2 ns. We observed a very quick convergence of the dynamical work for switching times of 0.05 ns. The Gibbs free-energy of the liquid phase at 2200~K as a function of pressure, as well as the evolution of the atomic volume with pressure are shown in Fig.~\ref{Fig:AS1}. Finally, we use the RS method to compute the Gibbs free-energy curve for the liquid phase as a function of temperature at 15.0 GPa using an identical procedure to that used in the case of Si-$\beta$-tin. This is achieved using the following fix commands in the \texttt{LAMMPS} script:

\texttt{variable lambda equal 1/(1+(elapsed/\$\{t\_s\})*(1/\$\{lf\}-1))}

\texttt{fix f1 all nph iso 15000.0 15000.0 1.0 ners v$\_$lambda}

The RS path chosen was such the scaling parameter $\lambda$ is varied between $\lambda=1$ to $\lambda=\lambda_{f}=2200/1000$. Thus, covering a temperature range between $T_{0}=2200$ K and $T_{f}=1000$ K. The temperature dependence of the Gibbs free-energy is then computed using Eq.~\ref{eqn:Gibbs_RS}, with a switching time of $t_{s}=0.5$ ns. The dynamical work was determined as the average over 10 independent realizations of the RS process along the forward and backward directions. Fig. \ref{fig:Tm2} shows the melting temperature of silicon  at 15.0 GPa at $T_{m}=1365.3(3)$ K.

\subsection{Pressure coexistence point between solids phases} 
\begin{figure}[htbp]
    \centering
         \includegraphics[scale=0.44]{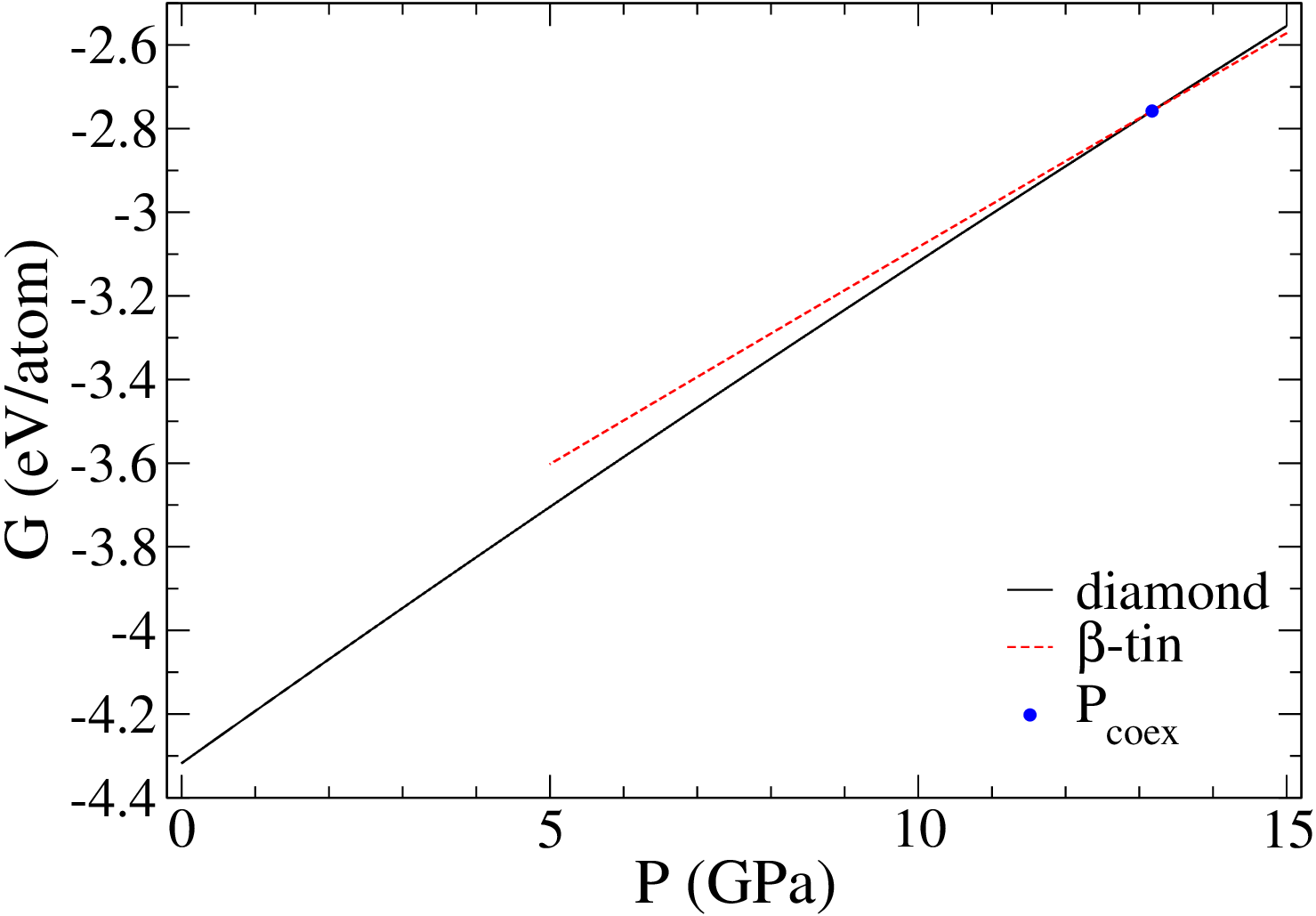}
     \caption{\label{fig:Pcoex} Gibbs free-energy per atom of Si-cd and Si-$\beta$-tin phases at 400 K. The crossing indicates the pressure coexistence at 400 K.}
\end{figure}

To determine the pressure coexistence of the solid phases at 400 K, we compute the Gibbs free-energy curves $G(P, T)$ for both solids as a function of pressure along an isothermal path by applying the AS method (Eq.~\ref{eqn:Gibbs_AS}). From the knowledge of the Gibbs free-energy for Si-cd at 400 K and zero pressure, we compute the dynamical work required to change the pressure dynamically between 0.0 GPa to 15.0 GPa, in order to determine the Gibbs free-energy of Si-cd at 15.0 GPa and 400 K. The system at both pressures was equilibrated during 0.1 ns prior to the switching runs. The unbiased dynamical estimators of the mechanical work were obtained from 10 independent forward and backward realizations using switching times between 0.01 ns to 0.2 ns. Also in this case, we observed a very quick convergence of the dynamical work for switching times of 0.05 ns. For Si-$\beta$-tin, the pressure is varied dynamically between 15.0 GPa to 5.0 GPa, using the same protocol used for Si-cd. Again is observed a very quick convergence of dynamical work for switching times of 0.05 ns. Fig. \ref{fig:Pcoex} shows the coexistence pressure at 400K of the solid phases at the value of $P_{coex}=13.17346(5)$ GPa, which is in agreement with the value previously reported by Romano et al. \cite{Romano2014}.
\subsection{Phase diagram of silicon:}
First, we determine the phase boundary between the Si-cd and liquid phases using the coexistence point determined at 0.0 GPa, as an initial condition to apply the dCCI method. In this case, the external pressure was chosen to be the independent variable in the integration process and varied linearly between the boundaries $P=0.0$ GPa and $P=10.2$ GPa for scaling times between 1 ps to 200 ps for testing the convergence of the calculation. Each process was started from equilibrated configurations in both cells corresponding to the known coexistence condition at $P=0.0$ GPa.

These calculations are accomplished using the following commands, \texttt{fix adapt/dcci} and \texttt{dcci} together, the command  \texttt{fix adapt/dcci} has a similar functionality of \texttt{fix adapt}, but in this case the scaling parameter is controlled by the \texttt{dcci} command, which must run using one or more processors per each phase. The usage of these commands in the \texttt{LAMMPS} script is:

\texttt{fix f1 all nph iso \$\{Pcoex\} \$\{Pcoex\} \$\{Pdamp\}}

\texttt{fix f2 all langevin \$\{Tcoex\} \$\{Tcoex\} \$\{Tdamp\} 1234 zero yes}

\texttt{fix f3 all adapt/dcci \$\{lambda\} pair sw fscale * *}

\texttt{dcci \$\{Tcoex\} \$\{Pcoex\} \$\{lambda\} f3 f1 \$\{t\_s\} press \$\{Pi\} \$\{Pf\}}

The \texttt{dcci} command performs the time-dependent calculation of the coexistence curve,  \texttt{\$\{Tcoex\}} and \texttt{\$\{Pcoex\}} define the initial coexistence condition, \texttt{\$\{lambda\}} is the scaling parameter and must be defined as well in the command \texttt{fix adapt/dcci}, which is identified by fix id \texttt{f3}. Also the command \texttt{dcci} needs to control the external pressure of the barostat, which is identified by the command fix id \texttt{f1}, \texttt{\$\{t\_s\}} is the scaling time. Finally the syntax \texttt{press \$\{Pi\} \$\{Pf\}}, which indicates the initial and final pressures along the coexistence the curve, is used to obtain the coexistence temperature as a function of the pressure i.e. $T_{coex}(P)$. 
\begin{figure}[tbp]
    \centering
         \includegraphics[scale=0.44]{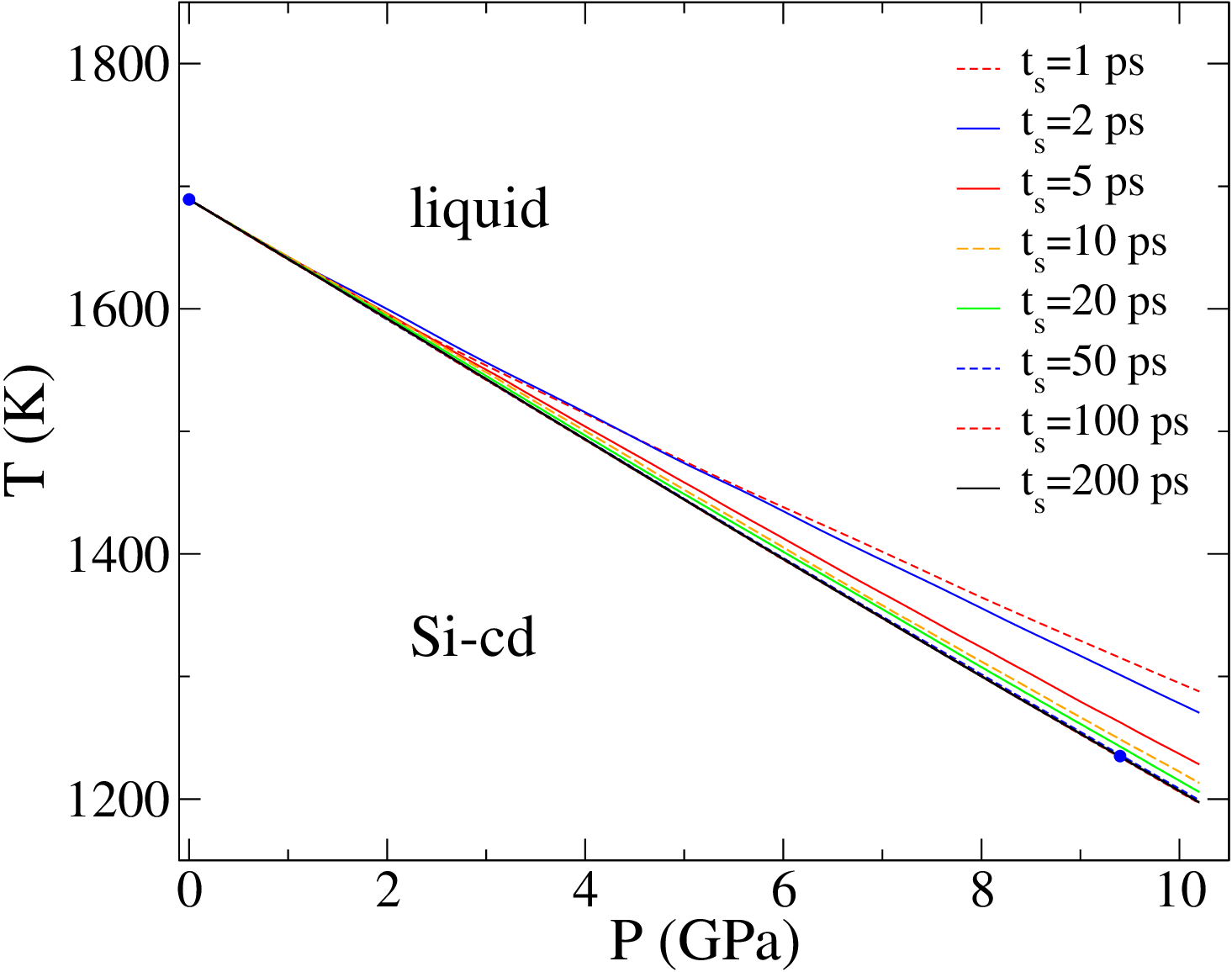}
     \caption{\label{fig:dcci1} Melting curves of Si-cd for different scaling times. The points represent the melting temperature at zero pressure and 9.4 GPa determined through the NE methods}
\end{figure}

In Fig. \ref{fig:dcci1} it is shown the phase boundaries between Si-cd and liquid for different scaling times. We observe the convergence of the results for scaling times above 50.0 ps. In order to estimate the effects associated with the irreversible nature of dCCI, we investigated the convergence of the coexistence temperatures as a function of the scaling time $t_{s}$, compared with the melting point of Si-cd at $P=9.4$ GPa calculated by the AS and RS methods (see Supplementary material Figs. S1-S4).

Fig. \ref{fig:dcci2} shows the convergence of the coexistence temperature at the pressure P=9.4 GPa determined from dCCI runs. The coexistence temperature converges to the RS result in approximately 50 ps per process, reaching a plateau above 100 ps.

\begin{figure}[tbp]
    \centering
         \includegraphics[scale=0.44]{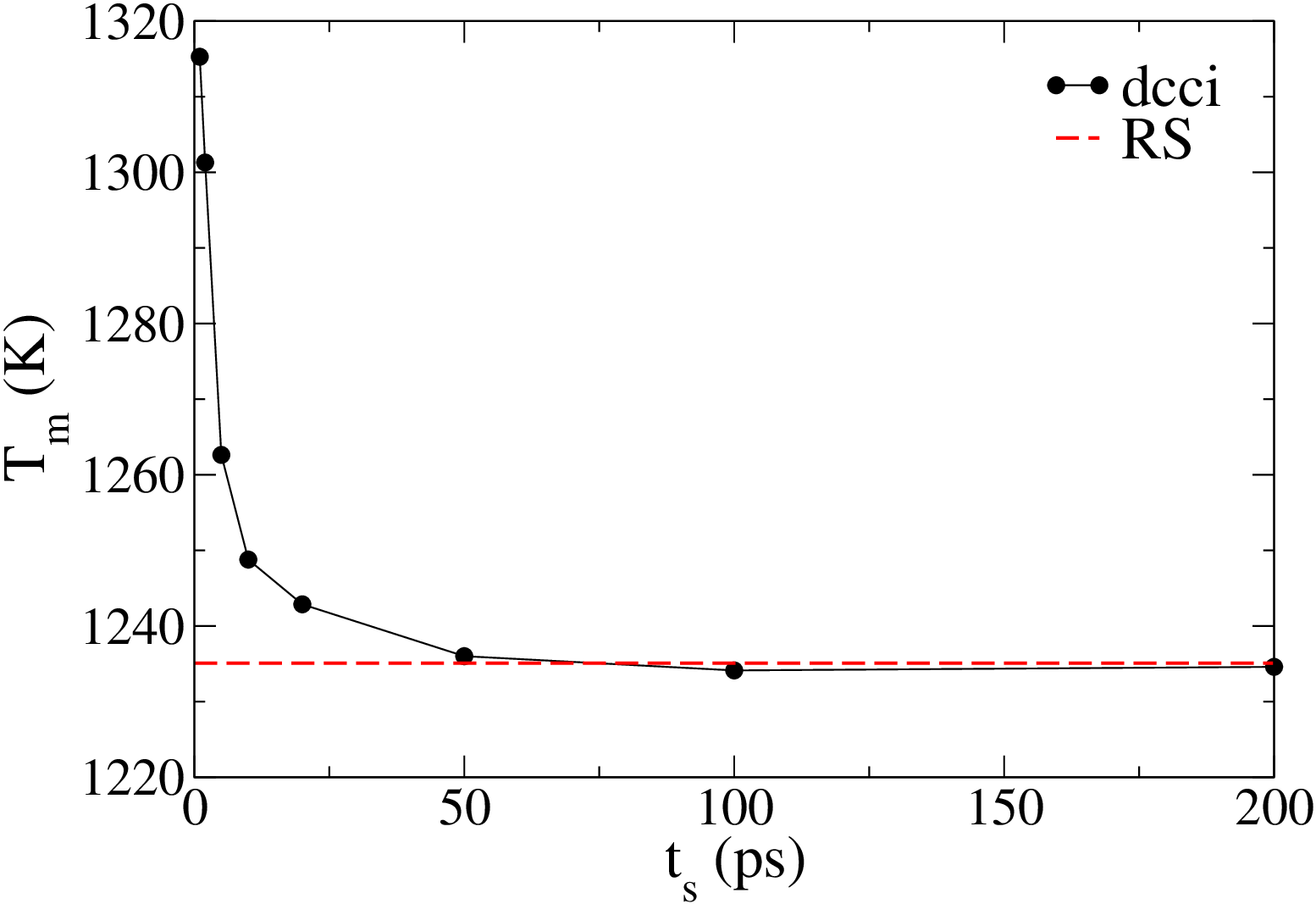}
     \caption{\label{fig:dcci2} Points indicate the melting temperature of Si-cd at P=9.4 GPa as a function of the scaling time determined by the dCCI method and the dashed red line corresponds to the value obtained by the RS method.}
\end{figure}

We now turn to the other phase boundary between the Si-$\beta$-tin and liquid phase. By using the coexistence point determined at 15.0 GPa, as an initial condition to apply the dCCI method, the pressure is varied linearly between the boundaries P=15.0 GPa and P=9.4 GPa for an scaling time of 100 ps. The procedure is practically identical to the previous case, but with an import difference, in the case of Si-cd and liquid phases, where we used isotropic volume fluctuations, because both phases present isotropic symmetry \texttt{fix f1 all nph iso}, in this case, the Si-$\beta$-tin phase presents anisotropic symmetry, to include the correct volume fluctuations for each system, i.e. isotropic for liquid \texttt{fix nph iso} and anisotropic \texttt{fix nph aniso} for the Si-$\beta$-tin, we need to define a different style of barostat for each phase in the \texttt{LAMMPS} script, this is achieved using the following commands:

\texttt{variable barostat word iso aniso}

\texttt{fix f1 all nph \$\{barostat\} \$\{Pcoex\} \$\{Pcoex\} \$\{Pdamp\}}

\texttt{fix f2 all langevin \$\{Tcoex\} \$\{Tcoex\} \$\{Tdamp\} 1234 zero yes}

\texttt{fix f3 all adapt/dcci \$\{lambda\} pair sw fscale * *}

\texttt{dcci \$\{Tcoex\} \$\{Pcoex\} \$\{lambda\} f3 f1 \$\{t\_s\} press \$\{Pi\} \$\{Pf\}}
\begin{figure}[t]
    \centering
         \includegraphics[scale=0.44]{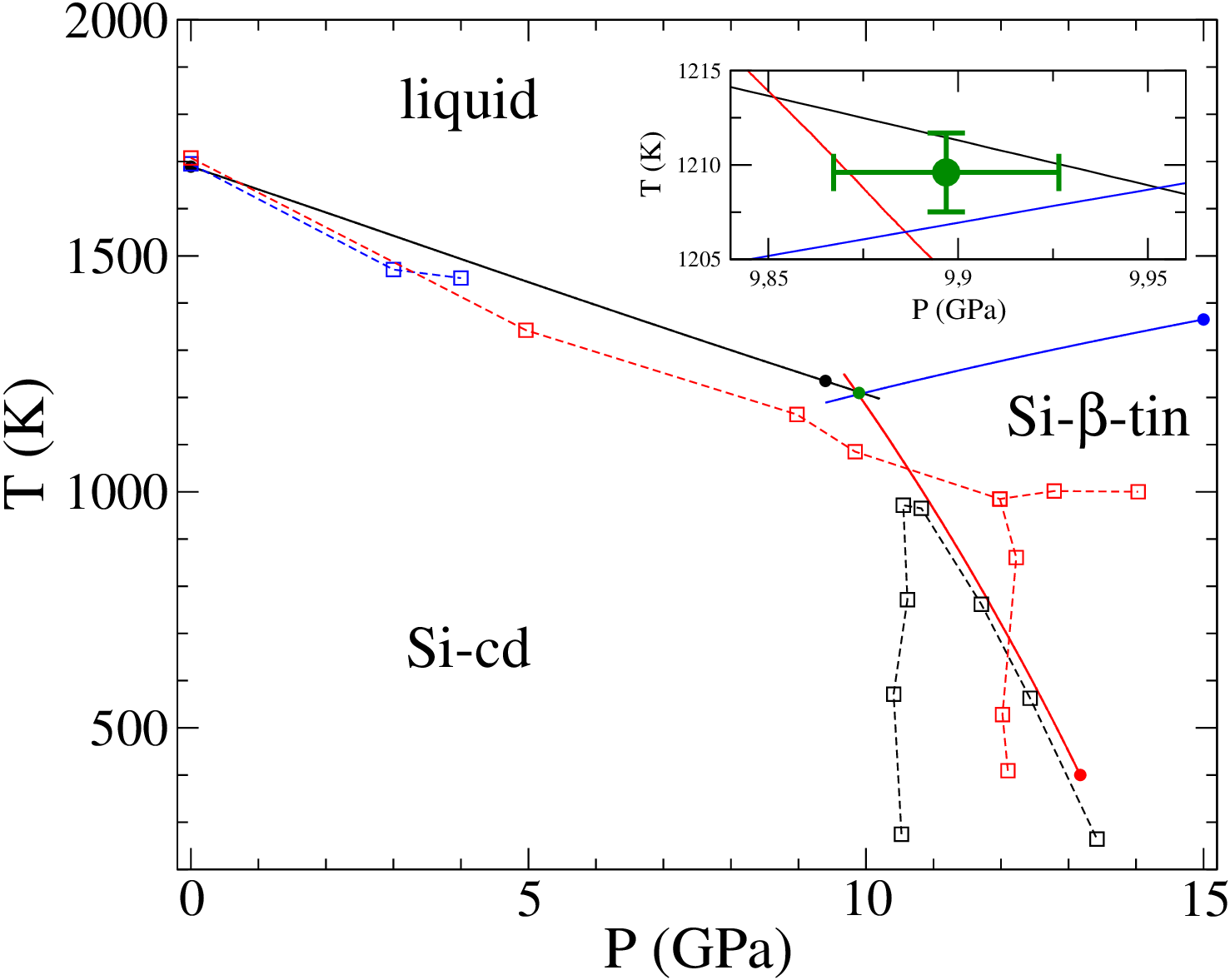}
     \caption{\label{fig:dcci3} Phase boundaries between silicon phases by the dCCI method. The dashed lines correspond to experimental results (the symbols {\color{blue}$\square$} Ref. \citenum{Jayaraman1963}, {\color{red}$\square$} Ref. \citenum{Bundy1964} and {\color{black}$\square$} Ref. \citenum{Voronin2003}). Symbols {\color{black}$\square$} to the left indicate the pressure in which for a given temperature the Si-$\beta$-tin phase begins to appear, whereas symbols {\color{black}$\square$} to the right indicate the pressure in which for that given temperature the Si-cd phase disappears completely.  \textbf{Inset}: triple point is indicated by the crossing of coexistence curves.}
\end{figure}
Next, we determine the coexistence curve between the solid phases, by using the coexistence point determined at 400 K, as an initial  condition to apply the dCCI method. In this case is more convenient to control the temperature, because the coexistence curve between the solid phases exhibit a very small change in pressure, whereas temperature varies in a large interval. Therefore, temperature is varied linearly between 400 K to 1250 K in a scaling time of 100 ps, each process was started from equilibrated configurations in both cells corresponding to the known coexistence condition at 400 K, with isotropic and anisotropic volume fluctuations for the Si-cd and Si-$\beta$-tin, respectively. This is achieved using the following commands:

\texttt{variable barostat word iso aniso}

\texttt{fix f1 all nph \$\{barostat\} \$\{Pcoex\} \$\{Pcoex\} \$\{Pdamp\}}

\texttt{fix f2 all langevin \$\{Tcoex\} \$\{Tcoex\} \$\{Tdamp\} 1234 zero yes}

\texttt{fix f3 all adapt/dcci \$\{lambda\} pair sw fscale * *}

\texttt{dcci \$\{Tcoex\} \$\{Pcoex\} \$\{lambda\} f3 f1 \$\{t\_s\} temp \$\{Ti\} \$\{Tf\}}
the syntax is similar to that described previously, but in this case we need to indicate the control of the temperature in the \texttt{dcci} command, using the following command \texttt{temp \$\{Ti\} \$\{Tf\}}, which indicates the initial and final temperature along the coexistence the curve to obtain the coexistence pressure as a function of temperature ($P_{coex}(T)$). 

Finally, in Fig. \ref{fig:dcci3} the phase diagram of silicon modeled by the SW potential is shown for the three phases considered in this work. The found triple point is $T_{p}$=(1210(2) K, 9.90(3) GPa), in agreement with previous values reported by Romano et al. \cite{Romano2014}. We have also compared our results with previous experimental results \cite{Jayaraman1963,Bundy1964,Voronin2003}. We note discrepancies between the phase diagram calculated using non-equilibrium methods and experimental results, can be attributed to the potential model chosen to describe the silicon phases, since originally the Stillinger-Weber potential was developed to study silicon in the diamond and liquid phase at low pressure, as can be observed by the agreement between our results and the experimental findings.

Despite the quantitative discrepancies between the computed and experimental phase diagrams, qualitative agreement between them is observed: such as the negative slope of the fusion curve of the silicon diamond phase, the positive slope of the fusion curve of the silicon $\beta$-tin phase, and the negative slope of the phase boundaries between the solids phases. Furthermore, it is remarkable that the crossing of the calculated coexistence lines define a very small region, spanning few degrees and hundredths of GPa, whose size is similar to that defined by the error bars in temperature and pressure. Thus, clearly defining the triple point for this model of Si. However, considering that the SW potential was designed to provide the experimental melting temperature of the Si-cd phase at low pressure and the description of other solid phases, such as Si-$\beta$-tin, were not taken into account, the agreement between our results and those from experiments is quite reasonable.
\section{Summary}\label{sec:conclusions}
This paper provides a guide for computing the Gibbs free-energy in a wide interval of the temperatures and pressures, which is used to determine the phase boundaries through the dCCI method within the \texttt{LAMMPS} MD simulation package. In addition, it describes implementation details in \texttt{LAMMPS} and makes available the computational tools in the form of full source code, scripts and auxiliary files.

As an illustrative example, the phase diagram of silicon model by SW potential was determined. Initially, the coexistence states between the phases of silicon were determined by the use of AS and RS methods, that allowed to efficiently calculate the Gibbs free-energy of the phases in a wide interval of temperatures and pressures. After that, with the knowledge the coexistence states between phases, the dCCI method was applied to determine the phase boundaries of Silicon using single non-equilibrium simulations. 

The techniques described in this paper, together with the supplied source code,  scripts  and  post-processing  files,  provide  a  platform  for  computing the phase boundaries of atomistic systems, which can be easily and efficiently determined using the \texttt{LAMMPS} software.

\section*{Acknowledgments} 
S.C. acknowledges the Brazilian agency FAPESP for the Post-Doctorate Scholarship under Grant No. 21/03224-3. 

A.A. gratefully acknowledges support from the Brazilian agencies CNPq and FAPESP under Grants No. 2010/16970-0, No. 2013/08293-7, No. 2015/26434-2, No. 2016/23891-6, No. 2017/26105-4, and No. 2019/26088-8. The calculations were performed at CCJDR-IFGW-UNICAMP and at CENAPAD-SP in Brazil.

\section*{Appendix A. Supplementary data} 
The following are the Supplementary data to this article can be found online at
\url{https://drive.google.com/file/d/1wcnm0V9Uun0ZDzjEfcHtn_D5eGmo2LZ9/view?usp=sharing}.
%
%

\bibliographystyle{elsarticle-num}



\begin{thebibliography}{10}
\expandafter\ifx\csname url\endcsname\relax
  \def\url#1{\texttt{#1}}\fi
\expandafter\ifx\csname urlprefix\endcsname\relax\def\urlprefix{URL }\fi
\expandafter\ifx\csname href\endcsname\relax
  \def\href#1#2{#2} \def\path#1{#1}\fi

\bibitem{Kirkwood1935}
J.~G. Kirkwood, Statistical mechanics of fluid mixtures, J. Chem. Phys. 3~(5)
  (1935) 300--313.

\bibitem{Freitas2016}
R.~Freitas, M.~Asta, M.~de~Koning, Nonequilibrium free-energy calculation of
  solids using lammps, Comput. Mater. Sci. 112 (2016) 333 -- 341.

\bibitem{Leite2016}
R.~Paula~Leite, R.~Freitas, R.~Azevedo, M.~de~Koning, The uhlenbeck-ford model:
  Exact virial coefficients and application as a reference system in
  fluid-phase free-energy calculations, J. Chem. Phys. 145~(19) (2016) 194101.

\bibitem{Leite2019}
R.~P. Leite, M.~de~Koning, Nonequilibrium free-energy calculations of fluids
  using lammps, Comput. Mater. Sci. 159 (2019) 316--326.

\bibitem{Cajahuaringa2018}
S.~Cajahuaringa, A.~Antonelli, Stochastic sampling of the isothermal-isobaric
  ensemble: Phase diagram of crystalline solids from molecular dynamics
  simulation, J. Chem. Phys. 149~(6) (2018) 064114.

\bibitem{Cajahuaringa2019}
S.~Cajahuaringa, A.~Antonelli, Nonequilibrium free-energy methods applied to
  magnetic systems: The degenerate ising model, J. Stat. Phys. 175 (2019)
  1572--9613.

\bibitem{LAMMPS}
S.~Plimpton, Fast parallel algorithms for short-range molecular dynamics, J.
  Comput. Phys. 117~(1) (1995) 1 -- 19.

\bibitem{deKoning2001}
M.~de~Koning, A.~Antonelli, S.~Yip, Single-simulation determination of phase
  boundaries: A dynamic clausius-clapeyron integration method, J. Chem. Phys.
  115~(24) (2001) 11025.

\bibitem{Stillinger1985}
F.~H. Stillinger, T.~A. Weber, Computer simulation of local order in condensed
  phases of silicon, Phys. Rev. B 31 (1985) 5262--5271.

\bibitem{Watanabe1990}
M.~Watanabe, W.~P. Reinhardt, Direct dynamical calculation of entropy and free
  energy by adiabatic switching, Phys. Rev. Lett. 65 (1990) 3301.

\bibitem{Frenkel}
D.~Frenkel, B.~Smit, Understanding molecular simulation: from algorithms to
  applications, 2nd Edition, Academic Press, 2002.

\bibitem{deKoning1997}
M.~de~Koning, A.~Antonelli, Adiabatic switching applied to realistic
  crystalline solids: Vacancy-formation free-energy in copper, Phys. Rev. B 55
  (1997) 735.

\bibitem{kofke1993}
D.~A. Kofke, Direct evaluation of phase coexistence by molecular simulation via
  integration along the saturation line, J. Chem. Phys. 98~(5) (1993)
  4149--4162.

\bibitem{Samuelgithub}
S.~Cajahuaringa, Nonequilibrium Free-energy calculation of Phase-Boundaries using \textsc{LAMMPS}, (2020) https://github.com/samuelcajahuaringa/dCCIforLAMMPS.

\bibitem{Parrinello1981}
M.~Parrinello, A.~Rahman, Polymorphic transitions in single crystals: A new
  molecular dynamics method, J. Appl. Phys. 52~(12) (1981) 7182--7190.

\bibitem{Schneider1978}
T.~Schneider, E.~Stoll, Molecular-dynamics study of a three-dimensional
  one-component model for distortive phase transitions, Phys. Rev. B 17 (1978)
  1302--1322.

\bibitem{Ryu2008}
S.~Ryu, W.~Cai, Comparison of thermal properties predicted by interatomic
  potential models, Model. Simul. Mater. Sci. Eng. 16~(8) (2008) 085005.

\bibitem{Romano2014}
F.~Romano, J.~Russo, H.~Tanaka, Novel stable crystalline phase for the stillinger-weber potential, Phys. Rev. B 90~(1) (2014) 014204.



\bibitem{Jayaraman1963}
A.~Jayaraman, W.~Klement,~Jr., G.~C.~Kennedy, Melting and Polymorphism at High Pressures in Some Group IV Elements and III-V Compounds with the Diamond/Zincblende Structure, Phys. Rev. 130~(540) (1963).

\bibitem{Bundy1964}
F.~P.~Bundy, Phase Diagrams of Silicon and Germanium to 200 kbar, 1000°C, J. Chem. Phys. 41~(12) (1964)
  3809--3814.

\bibitem{Voronin2003}
G.~A.~Voronin, C.~Pantea, T.~W.~Zerda, L.~Wang, Y.~Zha, In situ x-ray diffraction study of silicon at pressures up to 15.5 GPa and temperatures up to 1073 K, Phys. Rev. B 68~(R) (2003) 020102.

\end{thebibliography}





\end{document}